# Toward Diverse Polymer Property Prediction Using Transfer Learning

Elaheh Kazemi-Khasragh,[a,b], Carlos Gonzalez[a,b,c], Maciej Haranczyk[b,c]

[a]*Departamento de Ciencia de Materiales, Universidad Politécnica de Madrid, E.T.S. de Ingenieros de Caminos, Madrid, 28040, Spain*
[b]*IMDEA Materials Institute Getafe, Spain*
[c]*Corresponding authors emails: Carlos Gonzalez (carlosdaniel.gonzalez@imdea.org) and Maciej Haranczyk (maciej.haranczyk@imdea.org)*

## Abstract

The prediction of mechanical and thermal properties of polymers is a critical aspect for polymer development. Herein, we discuss the use of transfer learning approach to predict multiple properties of linear polymers. The neural network model is initially trained to predict the heat capacity in constant pressure ($C_p$) of linear polymers. Once, the pre-trained model is transferred to predict four additional properties of polymers: specific heat capacity ($C_v$), shear modulus, flexural stress strength at yield, and tensile creep compliance. They represent a diverse set of mechanical, thermal, and rheological proper- ties. We demonstrate the effectiveness of the approach by achieving high accuracy in predicting the four additional properties using relatively small datasets of 13 to 18 samples. Also, the performance of the base model is examined using five different loss functions. Our results suggest that the combined loss function had better performance compared to the individual loss functions.

## 1. Introduction

Synthetic polymers play a vital role in various industries and applications due to their abundant supply, diverse range of properties, and exceptional versatility. The demand for novel applications and improved efficiency has led to the exploration of new polymer materials, necessitating a thorough investigation of their properties and behaviors [2, 16].

Traditional modeling approaches, such as molecular dynamics, have limitations when it comes to predicting polymer properties. These approaches can be time-consuming, labor-intensive, and limited in their predictive accuracy. Furthermore, they often require extensive experimental validation data and may not capture the complex behavior and interactions at the molecular level that influence the properties of polymers. Moreover, polymers exhibit a wide range of structural and compositional variations, which further complicate the modeling process. The relationships between the molecular structure, processing conditions, and resulting properties are highly complicated and difficult to capture using traditional modeling methods alone. So, we need an alternative to overcome these limitations and explore new frontiers [24, 6, 32, 23].

Machine learning (ML) is a powerful technique that learn from data and make predictions or decisions without being explicitly programmed. Its ad- vantages lie in its ability to automate complex tasks, and uncover valuable insights from large datasets, driving innovation across industries [5, 18, 26, 13, 3, 22].

Besides of all ML advantages, the limited availability and variety of data pose a challenge to fully exploiting the potential of ML. Addressing the issue of insufficient data can be solved with the aim of transfer learning (TL). This approach encompasses various methodologies that involve re-purposing a model trained on one task for another related task. By leveraging pre- trained models, TL enables efficient knowledge transfer and enhances performance on new tasks with limited data availability [31, 29]. TL has emerged as a powerful technique which can be very helpful in the field of materials science [11, 14, 7]. However, this technique has been relatively under explored in the field of polymer characterization, with limited research studies available on this topic. Zhang et al. [35] utilize TL to successfully predict stress-strain curves of fiber reinforced polymer (FRP) composites fabricated via additive manufacturing. The results demonstrate the effectiveness of TL in achieving accurate predictions with limited training data, highlighting its potential in transforming the generation of stress-strain curves. Shi et al. [19] used TL technique to enhance the accuracy of ML models to predict adhesive free energy in polymer-surface interactions trained on small data sets. The TL approach significantly improves prediction accuracy and has implications for inverse materials design.

Yamada et al. [33] have developed some foundational models that serve as a basis for TL methods to predict material properties. They used these base models to predict the properties of inorganic crystalline materials, polymers, and small molecules. They examined these base models and transfer them to predict couple of properties with small size of dataset.

The focus of this work is to explore the use of transfer learning in a neural network model for efficient prediction of multiple properties of linear polymers. Initially, the model is trained on the heat capacity in constant pressure ($C_p$), a crucial property in polymer characterization with large dataset available. The transfer learning process involves adapting the pre- trained model (base model) to predict four additional diverse properties: specific heat capacity ($C_v$), shear modulus, flexural stress strength at yield, and tensile creep compliance. The objective of the study is to demonstrate the effectiveness of transfer learning by achieving high accuracy in predicting the four additional properties using relatively small datasets. The model's performance is evaluated using five different loss functions, emphasizing the importance of selecting an appropriate loss function for accurate predictions. Through this research, we show cast the benefits of transfer learning in efficiently predicting multiple properties of linear polymers, paving the way for advancements in various industrial and scientific applications.

## 2. Methods

### 2.1. Data collection

In this study, the data for heat capacity ($C_p$), specific heat capacity ($C_v$), flexural stress, shear modulus, and dynamic viscosity of polymers were collected from the PolyInfo database [25]. The descriptors for the prediction models were obtained from various sources including Alvadesc [20], Dragon [17], RDKit [27], and PaDEL2 [34]. Initially, a total of 14,321 descriptors were collected for the dataset. These descriptors provide valuable information about the chemical and structural characteristics of the polymers, which can be used to develop accurate prediction models for the target properties [15].

## 2.2. Principle component analysis

Principal component analysis (PCA) is a commonly employed technique the dimensionality reduction. It transforms a set of features into a set of uncorrelated principal components (PCs), which correspond to the directions in the feature space along which the data exhibits the most significant variation. One can then choose a subset of PCs that explain the desired threshold of the original data. The resulting principal components are orthogonal to each other and capture successively less variation in the data. One of the advantages of PCA is its ability to enhance the interpretability of the dataset. By reducing the dimensionality, the transformed data can be visualized and analyzed more easily. Additionally, employing PCA can lead to computational efficiency, as the reduced feature set requires less computational resources for model development [4, 12, 28].

## 2.3. Initial ML model

In this study, we employed a neural network model to predict the $C_p$ of 124 polymers. The neural network model was built using the Keras library with a sequential architecture. It consisted of 15 dense layers, with the ReLU activation function applied to the hidden layers. To optimize the model's performance, hyperparameter tuning was performed using Optuna, a hyperparameter optimization framework [1]. Optuna systematically explored to find best learning rate, regularization parameters, and the best combination for maximizing the model's accuracy. The hyperparameters such as epochs ranging from 500 to 1000 to strike a balance between under-fitting and over-fitting, the number of hidden layers varying between 5 and 20 for architectural complexity. Additionally, we dynamically determined the number of units in each layer within the range of 100 to 900, allowing flexibility in capturing complex patterns. The training dataset was randomly split into training and test sets, with a test size of 35%. The model was trained using the Adam optimizer and mean squared error (MSE) as the loss function. The performance of the model was evaluated using metrics such as mean squared error (MSE), squared correlation coefficient ($R^2$), and mean absolute error (MAE).

## 2.4. Loss function

In our study, the initial neural network (NN) model was formulated to predict $C_p$ with Mean Squared Error (MSE) serving as the primary loss function. To assess the impact of alternative loss functions on model performance, we experimented with five distinct functions: Mean Absolute Error (MAE), Huber Loss, Wing Shape Loss, and a combined loss. The choice of loss function plays a crucial role in assessing the discrepancy between the predicted output and the true target values [10]. MSE, and MAE are defined as in equation (1) and equation (2).

$$\text{MSE} = \frac{1}{N}\sum_{i=1}^{N}(y_i - \hat{y}_i)^2 \tag{1}$$

$$\text{MAE} = \frac{1}{N}\sum_{i=1}^{N}|y_i - \hat{y}_i| \tag{2}$$

We further employed the Huber loss function (HL) to model which is defined in equation (3):

$$\text{Huber Loss} = \frac{1}{N}\sum_{i=1}^{N}\begin{cases}\frac{1}{2}(y_i - \hat{y}_i)^2, & \text{if } |y_i - \hat{y}_i| \leq \delta \\ \delta\left(|y_i - \hat{y}_i| - \frac{1}{2}\delta\right), & \text{otherwise}\end{cases} \tag{3}$$

The loss is calculated as the average of the individual losses over all the samples. When the absolute difference between the true value $y_i$ and the predicted value $\hat{y}_i$ is less than or equal to the threshold $\delta$, the loss is computed as half of the squared difference. This corresponds to the squared loss (MSE) in this region. However, when the absolute difference exceeds the threshold, the loss is calculated as $\delta$ times the absolute difference minus half of $\delta$ squared. This introduces a linear loss instead of the quadratic loss, providing robustness to outliers. The choice of the threshold $\delta$ determines the transition point between the squared and linear loss regions and it is equal by one [30, 21].

The Wing Shape loss function (WSL) is a custom loss function used in machine learning models (equation 4). This function introduces two important parameters: $w$ and $\epsilon$. The parameter $w$ is a weight parameter that determines the threshold for the wing region. $\epsilon$ is a parameter that controls the smoothness of the loss function and can be adjusted based on the characteristics of the dataset to achieve the desired balance between robustness and sensitivity to deviations in the predictions. In this study $w$ and $\epsilon$ are 5 and 1.5, respectively [10, 9, 8].

$$\text{WSL} = \begin{cases} w \cdot \log\left(2 + \frac{|y_{true} - y_{pred}|}{\epsilon}\right), & \text{if } |y_{true} - y_{pred}| < w \\ |y_{true} - y_{pred}| - c, & \text{otherwise}\end{cases} \tag{4}$$

Where c calculated by:

$$c = 1.0 - \log\left(1.0 + \frac{w}{\epsilon}\right) \tag{5}$$

The combined loss function is a weighted sum of multiple loss functions which are discussed above. The formula for the combined loss function can be written as:

$$\text{Combined Loss Function} = \frac{1}{4}(\text{MSE} + \text{WSL} + \text{HL} + \text{MAE}) \tag{6}$$

By incorporating multiple loss functions, the combined loss function aims to leverage the strengths of each individual loss function and provide flexibility and, more comprehensive measure of the model's performance.

*2.5. Transfer learning*

After building the initial machine learning model to predict $C_p$, transfer learning was performed to further extend the predictive capabilities of the model. This method was used to predict other properties, namely $C_v$, flexural stress, shear modulus, and dynamic viscosity. The sizes of the respective datasets used for each property were 13, 13, 18, and 14 polymers, respectively. For the transfer learning process, the layers of the initial model were frozen and were transferred to the new models. By keeping the weights of the initial model unchanged, the focus was on training new layers dedicated to predicting the additional properties. Five new layers were added to the end of the transfer layers with weights in the range of 20-100. A low learning rate was utilized, allowing the model to gradually adapt to the new property prediction tasks while retaining the valuable knowledge acquired from the initial model.

These properties for prediction using transfer learning were selected based on their relevance and connection to the initial property, $C_p$, and their potential impact on material characterization. In addition to their relevance to the initial property, the chosen properties belong to different categories, namely mechanical, thermal, and rheological properties. This diverse selection was motivated by the interest in investigating the relationships and connections between these different aspects of the polymers' behavior. The mechanical properties, such as flexural stress and shear modulus, provide insights into how the polymers respond to applied forces and resist deformation. They are crucial for understanding the material's mechanical strength, structural integrity, and suitability for various applications. The thermal property, $C_v$, is related to the material's ability to store and release thermal energy. By incorporating $C_v$ we can gain a deeper understanding of the polymers' thermal behavior, their ability to absorb and dissipate heat, and their thermal stability. The rheological property, dynamic viscosity, characterizes the material's flow behavior and resistance to deformation under shear stress. It plays a significant role in processes such as polymer processing, coating, and flow behavior analysis. Including dynamic viscosity as an additional property allows for a comprehensive exploration of the polymers' rheological properties and their flow characteristics under different conditions. By considering properties from different categories, we aim to capture a broader spectrum of the polymers' characteristics and gain a more comprehensive understanding of their overall behavior. This multi-faceted analysis allows us to explore the relationships between mechanical, thermal, and rheological properties and their influence on each other, contributing to a more holistic characterization of the polymers.

## 3. Results and Discussion

### *3.1. Principle component analysis and base ML model*

Figure 1 demonstrates the explained variance ratio for 20 principal components (PCs). In the dataset, there are a total of more data points, each representing a principal component. However, for the purpose of visualization, only 20 principal components are shown in the figure depicting the explained variance ratio versus the number of principal components. Subsequently, a subset of 13 principal components was selected based on their cumulative explained variance ratios. Specifically, the cumulative explained variance ratios reached a threshold of approximately 0.7, indicating that these 13 principal components capture a significant portion of the overall variance in the dataset.

By examining the top contributing descriptors within the principal components, we uncovered details about their makeup. These descriptors of- fer a nuanced understanding of the molecular intricacies within the polymers. These descriptors highlight specific patterns, chemical structures, and how molecules are connected. For instance, Dragon7 descriptors, like B02[O-O], unravel specific structural patterns, potentially related to oxygen-containing motifs. Other descriptors, such as PubchemFP582 and PubchemFP11, provide binary representations of chemical substructures based on PubChem data. RDKit descriptors, exemplified by morgan363 and maccs157, capture molecular arrangements and specific features. Additionally, topological descriptors like topological929 and topological308 shed light on the connectivity and spatial arrangement of atoms within the molecules. This comprehensive exploration of diverse descriptors enriches our grasp of the molecular composition and structural nuances inherent in the polymers under investigation and providing essential insights into the structural characteristics influencing the thermal and mechanical behavior of the polymers.

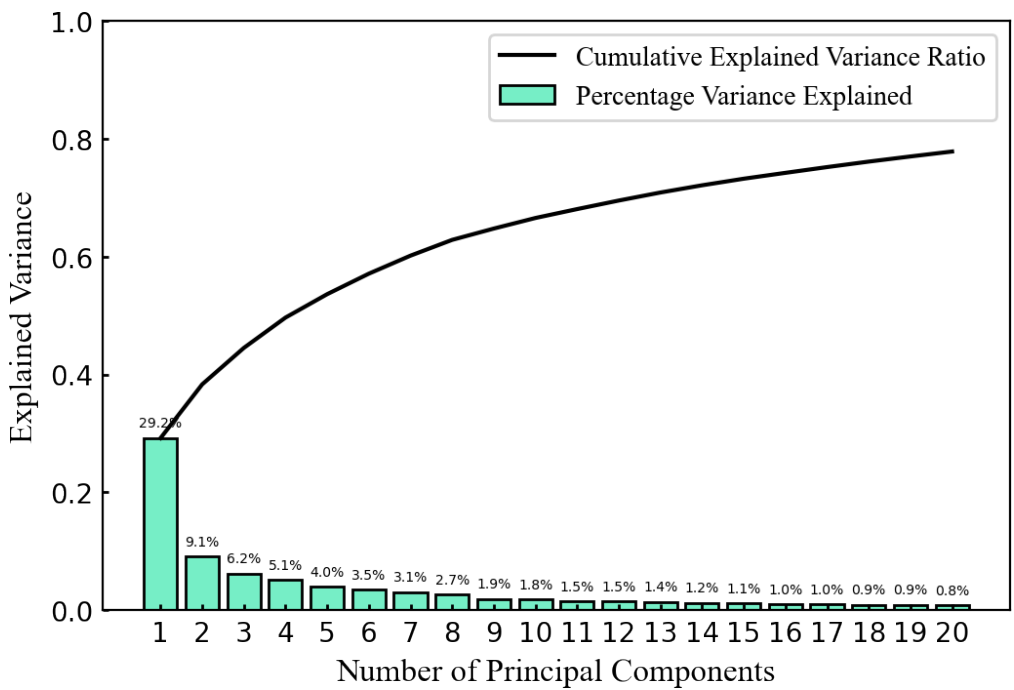

**Figure 1**. Explained variance ratio for 20 principal components.

The polymer chemical space is illustrated in Figure 2 using principal components obtained through PCA on the descriptors of all polymers. The first four principal components (PC1, PC2, PC3, and PC4) reveal the intricate relationships between polymers coming from various sets employed to predict different output properties, represented by distinct colors in the plot. Specifically, in Figure 2(a) through(d) respectively represent the relationships between PC1 and PC2, PC1 and PC3, PC1 and PC4, and PC2 and PC3.

The spread of the points in the plot suggests that, although differentiated by size, the datasets roughly occupy the same area of the chemical space. The absence of clustering or separation indicates that the descriptors used in the transfer learning model effectively captured common underlying fac- tors influencing these properties. This observation supports the suitability of the transfer learning approach for predicting multiple properties simultaneously. The descriptors exhibit a degree of shared information among different properties, contributing to the comprehensive understanding of the polymer system. By utilizing the transfer learning model trained on the initial $C_p$ prediction task and applying it to predict additional properties, we leverage the knowledge gained from the initial training and extend it to other related properties. This approach offers the advantage of exploiting the shared information and relationships among the properties, leading to improved pre- dictions and a more comprehensive understanding of the polymer system.

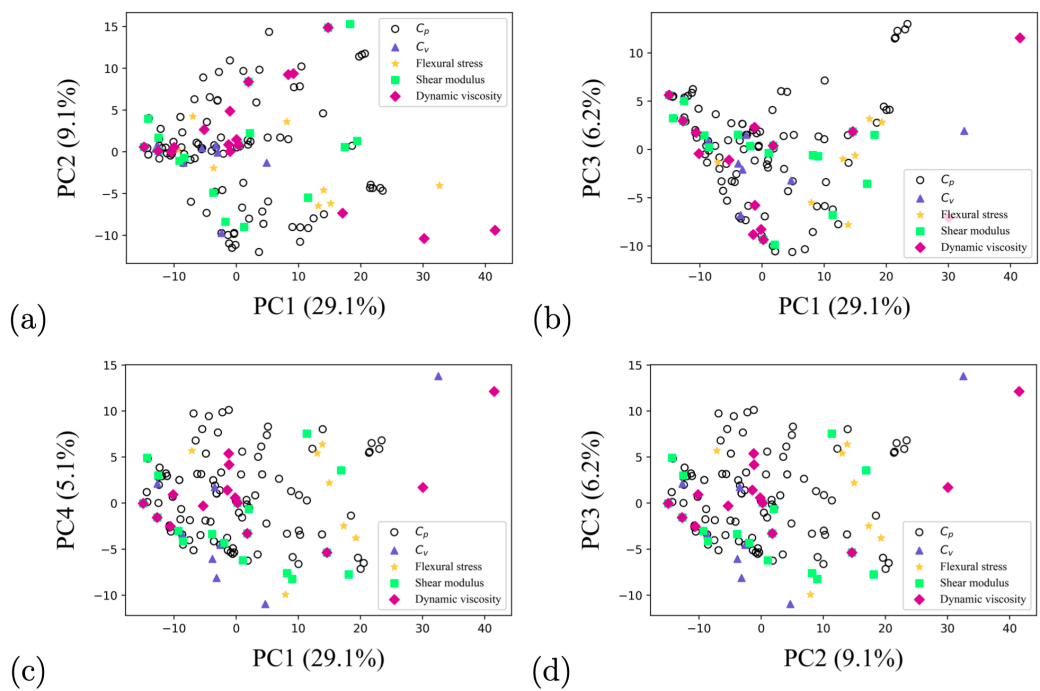

**Figure 2**. Comprehensive principal component analysis depicting relationships for predicting $C_p$, $C_v$, flexural stress, shear modulus, dynamic viscosity (a) PC1 vs. PC2, (b) PC1 vs. PC3, (c) PC1 vs. PC4, and (d) PC2 vs. PC3.

In figure3, we present a comparison between the expected and predicted values of the neural network (NN) model for the $C_p$ of polymers. The close alignment of the data points with a small deviation from the diagonal line indicates that the model performs well in predicting $C_p$. The proximity of the points to the diagonal line demonstrates the accuracy and precision of the model's predictions. These results prove the effectiveness of the NN model in estimating the heat capacity of polymers based on the selected descriptors.

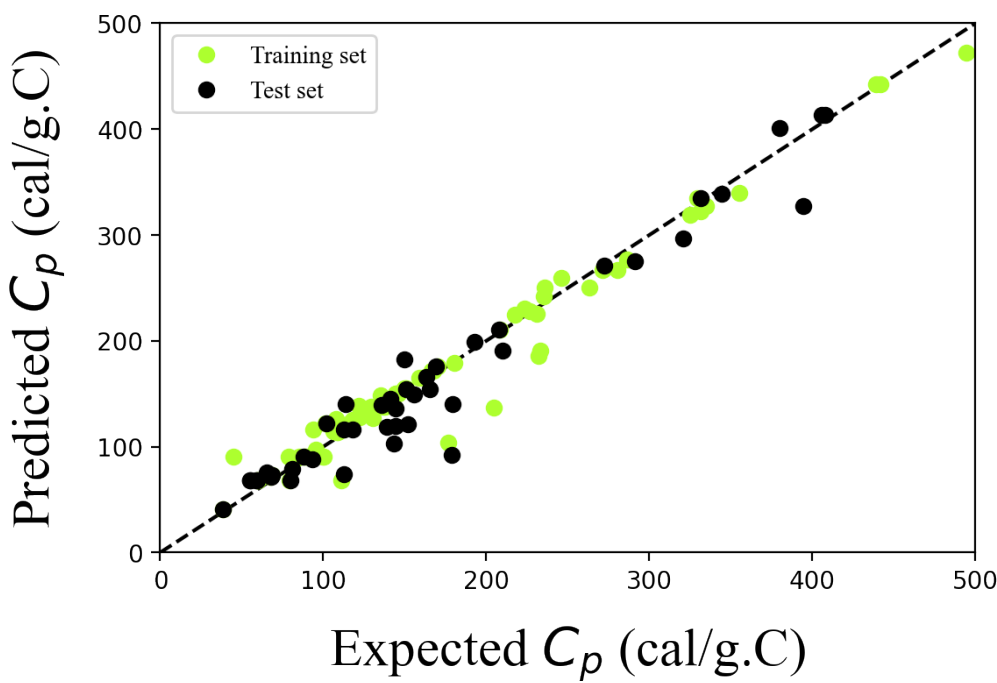

**Figure 3**. A comparison between the expected and predicted values of the NN model for $C_p$.

*3.2. Loss Function*

The accuracy of the NN models constructed with five different loss functions (MSE, MAE, Huber Loss, wing shape loss, and combined loss), in the $C_p$ are gathered in Table 1. Notably, it is crucial to consider the consistent units for $C_p$, which are expressed in calories per gram per degree Celsius (cal/g°C), when interpreting the values of the loss functions presented in Table 1.

When comparing the $R^2$ values, we observe that all the loss functions yield high values, indicating a good fit of the model to the training and testing data. The MSE loss function achieved an $R^2$ value of 0.962 on the training set and 0.93 on the testing set. Similarly, the MAE, Huber Loss, and Wing Shape Loss functions achieved $R^2$ values of 0.96, 0.967, and 0.967 on the training set, and 0.948 on the testing set.

Also, the MSE and MAE statistics are used to evaluate the performance of models with different loss function. The model utilizing the combined loss function showcased the most favorable results, yielding the lowest MSE (12.42) and MAE (15.86) values. Nevertheless, the Huber Loss and Wing Shape Loss functions displayed competitive performance as well, with MSE values of 12.47 and 12.32, and MAE values of 15.82 and 15.49, respectively. In the case of the model with MSE loss function the results are higher than others model.

Interestingly, the combined loss function, which incorporates a balanced combination of the MSE, MAE, Huber Loss, and Wing Shape Loss, resulted in the highest $R^2$ values of 0.97 on the training set and 0.95 on the testing set. This indicates that the combined loss function effectively captures the strengths of the individual loss functions, resulting in improved prediction accuracy.

Overall, based on the evaluation metrics and the results presented in Table 1, it can be concluded that the combined loss function outperforms the other individual loss functions in predicting the $C_p$. The combined loss function provides a robust and balanced approach, considering multiple aspects of the data, and yields more accurate predictions.

**Table 1.** Accuracy results of the NN trained model in predicting $C_p$ with different loss function

| Loss function | $R^2$ _ train | $R^2$ _ test | MSE | MAE |
|---|---|---|---|---|
| MSE | 0.962 | 0.933 | 16.10 | 19.07 |
| MAE | 0.96 | 0.948 | 12.42 | 15.86 |
| Huber loss | 0.967 | 0.948 | 12.47 | 15.82 |
| wing shape | 0.967 | 0.948 | 12.32 | 15.49 |
| Combined | 0.971 | 0.95 | 12.1 | 15.2 |

*3.3. Transfer Learning*

Once the base model to predict $C_p$ has been built and the performance of the model is approved in the previous section. In this section, we will discuss the effective utilization of transfer learning in four distinct scenarios, showcasing its successful applications. The visual representation in Figure 4 provides evidence of the transferability between $C_v$, flexural stress, shear modulus, dynamic viscosity, and $C_p$. The transfer learning models demonstrate strong predictive capabilities, as evident from the expected versus predicted values depicted in Figure 4. The high accuracy achieved in predicting $C_v$, flexural stress, shear modulus, dynamic viscosity, and $C_p$ further sup- ports the effectiveness of the transfer learning approach. The performance metrics of the models are summarized in Table 2, highlighting their successful predictive performance. The high value of the $R^2$ low MAE and the correspond to the high accuracy of the model. $R^2$ of the train set ranges from
0.98 to 0.91, and $R^2$ of the test set ranges from 0.89 to 0.83. The model for the dynamic viscosity shows the lower value of $R^2$ of the test set, 0.83 for the given property. On the contrary, the model for the $C_v$, performs better than the other models, with the low mean absolute error and $R^2$ of the test set of 0.89.

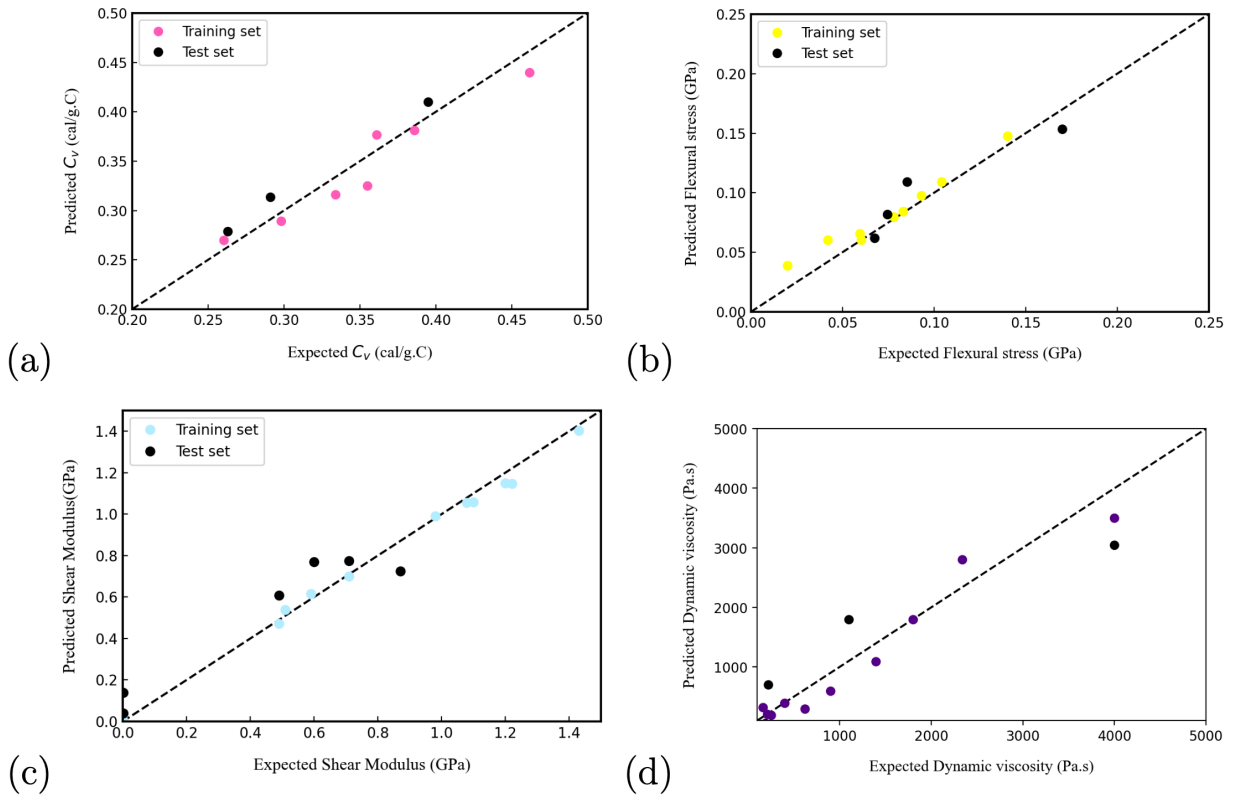

**Figure 4**. Expected and predicted values of the model to predict (a) $C_v$, (b) flexural stress, (c) shear modulus, and (d) dynamic viscosity.

**Table 2.** Performance of ML models built with transfer learning

| Models | $R^2$_train | $R^2$_test | MSE |
|---|---|---|---|
| $C_v$ | 0.91 | 0.90 | $1.1 \times 10^{-4}$ |
| Flexural Stress | 0.91 | 0.86 | $5.8 \times 10^{-5}$ |
| Shear Modulus | 0.98 | 0.87 | $2.4 \times 10^{-2}$ |
| Dynamic Viscosity | 0.94 | 0.83 | $10 \times 10^{4}$ |

## 4. Conclusion

In conclusion, this study employed a comprehensive approach to predict multiple polymer properties. We first conducted PCA analysis to reduce the dimensionality of a dataset representing ca. 150 materials represented by 27890 descriptors. By considering the cumulative explained variance ratio a subset of 13 principal components was selected. Then the initial NN model utilized 13 PCs was constructed to predict $C_p$ of polymers and the model exhibited good accuracy.

Furthermore, we explored the performance of the NN model using five different loss functions: MSE, MAE, Huber Loss, Wing Shape Loss, and a combined loss function. Our findings indicate that the combined loss function outperformed the individual loss functions, highlighting the advantage of incorporating multiple criteria in the prediction process.

Additionally, we successfully applied transfer learning by freezing the layers of the initial NN model and transferring the knowledge to predict $C_v$, flexural stress, shear modulus, and dynamic viscosity. Through PCA analysis, we observed correlations and relationships among the predicted properties, suggesting the presence of shared underlying factors captured by the selected descriptors. This supports the suitability of TL for simultaneous prediction of multiple properties and emphasizes the relevance of these properties to one another.

In conclusion, our approach successfully leveraged PCA, NN modeling, and transfer learning to predict a diverse range of polymer properties. The combined loss function enhanced prediction accuracy, while the interrelated nature of the selected properties further facilitated successful transfer learning. By covering properties from different categories, our study offers valuable insights into the comprehensive characterization of polymers.

[32] William WL Wong and Forbes J Burkowski. "A constructive approach for discovering new drug leads: Using a kernel methodology for the inverse-QSAR problem". In: *Journal of cheminformatics* 1 (2009), pp. 1–27.

[33] Hironao Yamada et al. "Predicting materials properties with little data using shotgun transfer learning". In: *ACS central science* 5.10 (2019), pp. 1717–1730.

[34] Chun Wei Yap. "PaDEL-descriptor: An open source software to calculate molecular descriptors and fingerprints". In: *Journal of computational chemistry* 32.7 (2011), pp. 1466–1474.

[35] Ziyang Zhang, Qingyang Liu, and Dazhong Wu. "Predicting stress–strain curves using transfer learning: Knowledge transfer across polymer composites". In: *Materials & Design* 218 (2022), p. 110700.18